\documentclass[journal]{IEEEtran}

\usepackage{url}
\usepackage{graphicx}
\usepackage{balance}
\usepackage{graphics}
\usepackage{graphicx}
\usepackage{epsfig,amssymb}
\usepackage{epstopdf}
\usepackage{times}
\usepackage{mathrsfs}
\usepackage{array}
\usepackage{amsmath, latexsym, amsfonts, amssymb,bm}
\usepackage[mathscr]{eucal}
\usepackage{array, tabularx}
\usepackage[table]{xcolor}
\usepackage{multirow}
\usepackage{epsfig}
\usepackage{cite}
\usepackage{tikz}
\usepackage{mathtools}
\usepackage{psfragx}
\usepackage{xcolor}
\usepackage{bbm}
\usepackage{lineno}
\usepackage{algorithm}
\usepackage{algorithmic}
\usepackage{lipsum}
\usepackage{caption}
\usepackage{subcaption}
\usepackage{authblk}
\usepackage{etoolbox}

\newcommand{\defeq}{\ensuremath{\triangleq}} 
\newcolumntype{S}{>{\centering\arraybackslash}p{5em}}

\newtheorem{lemma}{Lemma}

\newtheorem{proposition}{Proposition}

\DeclareMathOperator{\Tr}{Tr}
\DeclareMathOperator{\rank}{rank}
\DeclareMathOperator{\diag}{{diag}}

\IEEEoverridecommandlockouts

\begin{document}
	
	\title{Channel Estimation  for Intelligent Reflecting
		Surface Assisted  Backscatter Communication}
	
	\author{ 
		Samith Abeywickrama, Changsheng You, \IEEEmembership{Member, IEEE},  Rui Zhang,  \IEEEmembership{Fellow, IEEE},  \\ and  Chau Yuen,  \IEEEmembership{Fellow, IEEE} 
		
		\thanks{			
			S. Abeywickrama is with the Department of Electrical and Computer Engineering, National University of Singapore, Singapore, and also with Singapore University of Technology and Design, Singapore (e-mail: samith@u.nus.edu).
			
			C. You and R. Zhang are with the Department of Electrical and Computer Engineering, National University of Singapore, Singapore (e-mail: \{eleyouc,elezhang\}@nus.edu.sg).
			
			C. Yuen is with  Singapore University of Technology
			and Design, Singapore (e-mail: yuenchau@sutd.edu.sg).
		}
	}
	
	\maketitle 
	
	\begin{abstract}
		Intelligent reflecting surface (IRS) is a promising technology to improve the  performance of backscatter communication systems by smartly reconfiguring the multi-reflection channel. To fully exploit the passive beamforming gain of IRS in backscatter communication, channel state information (CSI) is indispensable but more  practically challenging to acquire than conventional IRS-assisted systems, since IRS passively reflects signals over both the forward and backward (backscattering) links between the reader and 
		tag. To address this issue, we propose in this letter a new and efficient channel estimation scheme for the IRS-assisted backscatter communication system. To minimize the mean-square error (MSE) of channel estimation, we  formulate  and solve an optimization problem by designing the IRS training  reflection matrix for channel estimation under the constraints of unit-modulus elements  and full rank.  Simulation results verify the effectiveness of the proposed channel estimation scheme as compared to other baseline schemes.
	\end{abstract}

	\begin{IEEEkeywords}
		Intelligent reflecting surface, backscatter communication, channel estimation.
	\end{IEEEkeywords}

	\IEEEpeerreviewmaketitle
	
	\section{Introduction} 
	
	
	
	Intelligent reflecting surface (IRS) has recently emerged as a 	promising technology to enhance the spectrum and energy efficiency of future wireless systems cost-effectively \cite{wu2020intelligent}. Specifically, IRS is able to engineer favourable  wireless propagation environment via controlling signal reflection at its large number of passive reflecting elements. This thus has motivated active research recently in applying IRS to existing wireless systems, such as UAV communication, wireless power transfer, millimeter wave communication, mobile edge computing, and  backscatter communication \cite{wu2020intelligent,chongwang,9351782,9279326,9017956,jia2020intelligentxx}.

	Particularly, for IRS-assisted backscatter communication systems,  IRS can be properly deployed to enhance the channel gain over both the forward (transmission) and backward (backscattering) links between the reader and tag, where the reader transmits the carrier signal to the tag, while the tag appends its own information to the received signal and backscatters it to the reader.  However, to achieve the enormous passive beamforming gain brought by IRS, channel state information (CSI) is crucial and needs to be acquired, which, however, is more  practically challenging than conventional IRS-assisted systems. This is because both the IRS and tag can reflect signals only without sophisticated signal processing capability. Thus, only the composite channel, which is the product of the forward and backward channels, can be estimated at the reader based on the pilot signal sent by  itself. In particular, each of the forward/backward channel is the superposition of the reader-tag direct channel and the reader-IRS-tag cascaded channel. This makes the conventional channel estimation schemes for backscatter communication systems without IRS \cite{8618337} and IRS-assisted  communication systems without tag \cite{91331422,9195133,9053695,9129778,chongwangxxxx} inapplicable, thus motivating the current work to tackle this challenge. 
	
	\begin{figure}[t]
		\centering 
		\includegraphics[width=0.85\linewidth]{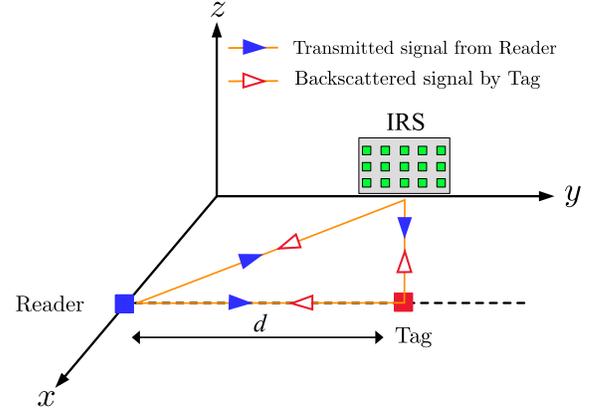}
		\caption{{IRS-assisted backscatter communication system.}}  
		\label{sm2} 
	\end{figure}	

	Specifically, we consider in this letter an IRS-assisted monostatic backscatter communication system as 	illustrated in Fig. \ref{sm2}, where an IRS is employed to assist the communication between a reader and a tag, both equipped with a single antenna.  A novel channel estimation scheme is proposed to 	efficiently estimate both the reader-tag direct channel and reader-IRS-tag reflecting channel, by controlling the IRS training reflections over time. Moreover, the IRS training matrix is  optimized to minimize the channel estimation error, for which the optimal solution is derived in closed form. Simulation results verify the effectiveness of the proposed channel estimation scheme as compared to other baseline schemes.


	\textit{Notations:} In this letter, vectors and matrices are denoted by bold-face lower-case and upper-case letters, respectively. For a complex-valued vector $ \mathbf v $, we denote by $ \| \mathbf v \| $, $ \mathbf v^T $, $ \mathbf v^H $,  $\diag(\mathbf v)$,  and $[{\mathbf v}]_{a:b}$ as its $\ell_2$-norm, transpose, conjugate transpose, a diagonal matrix with  diagonal elements being the corresponding elements in $ \mathbf v $,  and the elements from the $a$-th entry to the $b$-th entry  of $ \mathbf v $,  respectively. Scalar $ v_i$ denotes the $i$-th element of a vector $\mathbf v$. For a square matrix $ \mathbf{S} $, $ \Tr(\mathbf{S}) $ and $ \mathbf{S}^{-1} $ denote its trace and inverse, respectively. For any  matrix $ \mathbf{A} $, we denote by  $ \mathbf{A}^H $, $ \rank(\mathbf{A}) $, and $\mathbf{A}_{n,k}$ as its conjugate transpose, rank, and $(n, k)$th element, respectively. $ \mathbf{I} $  denotes an identity matrix  with appropriate dimensions. $ \mathbb{C}^{x \times y} $  denotes the space of $ x \times y $ complex-valued matrices. $ \jmath $ denotes the imaginary unit, i.e., $ \jmath^2 = -1 $.  For a complex-valued scalar $ \mathrm v $, we denote by $ | \mathrm v | $, $\arg(\mathrm v)$, and $  {\mathrm v}^\dag$  its absolute value, phase, and complex conjugate, respectively.  $|\Omega|$ denotes the cardinality of the set $\Omega$. $\mathbb{E}(\cdot)$ denotes the statistical expectation. $\otimes$ denotes the Kronecker product.

	\section{System Model}

	As shown in Fig. \ref{sm2}, we consider an IRS-assisted monostatic backscatter communication system, where an IRS composed of $N$ subsurfaces is deployed to assist in the   communication between a full-duplex (FD) single-antenna reader and a single-antenna tag. Let $ h_{d} \in \mathbb C$, $\mathbf f \in \mathbb C^{N\times 1}$, and  $\mathbf h_{r} \in \mathbb C^{N\times1}$ denote the baseband equivalent channels from the reader to tag, reader to IRS, and IRS to tag, respectively, which are assumed to remain constant within each channel coherence interval.	For the IRS, we denote by $\mathbf v=[\beta_{1} e^{\jmath\theta_{1}},\dots,\beta_{N} e^{\jmath\theta_{N}}]^H  \in \mathbb C^{N\times1}$  its  reflection  vector, where $\beta_{n}\in [0,1]$ and $\theta_{n}\in [0,2\pi)$, with  $n\in \{1,\dots,N\}$, respectively denote the (common) reflection amplitude and  phase shift of all  reflecting elements in   subsurface $n$. For simplicity, we set $\beta_{n}=1$ (or its maximum value), $\forall n$,  to maximize the IRS reflected signal  power in both channel training and data transmission.	

	Let $s $ denote the signal sent by the reader with power $P_t$. As no signal processing is performed at the tag and thus no noise added, the signal received at the tag over both the direct and IRS reflecting links is thus given by
	\begin{align} 
	y_{T} &= \big( h_{d} + \mathbf v^H \diag(\mathbf h_{r})\mathbf f   \big) s \nonumber \\
	&=( h_{d} + \mathbf v^H \mathbf h_c   ) s, \label{eq1}
	\end{align}
	where  $\mathbf h_c=\diag(\mathbf h_{r})\mathbf f \in \mathbb C^{N\times1}$ denotes  the cascaded reader-IRS-tag channel. Let $\alpha\in[0,1]$ denote the portion of signal reflected by the tag, which is simply set as $\alpha=1$ in the sequel without loss of generality. Then, based on channel reciprocity, the received signal at the reader is given by\footnote{In practice, a decoupler is usually integrated into the reader to enable the FD operation by suppressing the self-interference caused by the transmitted signal $s$ \cite{5467234}.}
	\begin{align} 
	y_{R} &= \alpha y_{T}\Big( h_{d} + \mathbf v^H \mathbf h_c     \Big) + z \nonumber\\
	&= \Big( h_{d} + \mathbf v^H \mathbf h_c     \Big)^2 s + z, \label{y_r}
	\end{align}
	where $z$ denotes the additive white Gaussian (AWGN) noise at the reader with power $\sigma^{2}$. For convenience, we normalize the noise power by $P_t$ and thereby assume $s=1$ without loss of generality; thus the symbol $s$ can be omitted in \eqref{y_r} in the sequel. We consider a two-phase data transmission protocol, where each channel coherence block of $T$ symbols is divided into two phases. During the first channel training phase, the reader consecutively transmits $T_t$ pilot symbols, while the  IRS properly sets its training reflections over time to facilitate the channel estimation at the reader. Based on the estimated channel,  the reader designs the IRS passive beamforming for data transmission, denoted by $\mathbf v^\ast$, and  sends it  to the IRS controller via a separate link for it to  tune the IRS reflection accordingly for aiding the tag's backscatter communication over the remaining $T-T_t$ symbols in the second data transmission  phase. From \eqref{y_r} and under the assumption of perfect CSI at the reader, the optimal IRS passive beamforming that maximizes the received signal-to-noise ratio (SNR) at the reader is given by 
	\begin{align} 
	\mathbf v^\ast = e^{\jmath\arg\big(  { h}_d^\dag \mathbf h_c\big)}. \label{opt}
	\end{align}
	
	\section{Proposed Channel Estimation and Training Design}
	
	
	It is observed from \eqref{opt} that the optimal IRS passive beamforming design for data transmission requires the CSI of both the reader-tag direct link and the  cascaded IRS reflecting link, i.e., ${ h}_d$ and $\mathbf h_c$, with totally $N+1$ complex-valued  channel coefficients. However, such  CSI cannot be  obtained by applying the existing IRS channel estimation methods (see e.g., \cite{9195133,9053695,91331422}) for the linear channel model with $N+1$ training symbols only, since $y_{R}$ in \eqref{y_r} is a nonlinear function of ${ h}_d$ and $\mathbf h_c$, leading to  the issue of \emph{sign ambiguity}. Specifically, for each training symbol $k$, the corresponding effective channel (assuming $z=0$ for the purpose of illustration) from the reader to tag is given by 
	\begin{align} 
	b_k \defeq { h}_d + \mathbf v_k^H\mathbf h_c = \pm \sqrt{y_{R}}, \label{ehau}
	\end{align}
	where $\mathbf v_k$ denotes the IRS reflection vector for the $k$-th symbol. Due to  the sign ambiguity in \eqref{ehau} for determining ${ h}_d + \mathbf v_k\mathbf h_c$, the conventional methods in \cite{9195133,9053695,91331422} with time-varying $\mathbf v_k$'s cannot be  applied to estimate ${ h}_d$ and $\mathbf h_c$ in the nonlinear channel model in \eqref{y_r} uniquely.

	\subsection{Proposed Channel Estimation Scheme} \label{formulation}
	
	We first present two useful lemmas as follows. 
	\begin{lemma} \label{lemma3}
		The received signal at the reader  in \eqref{y_r} can be equivalently  rewritten as
		\begin{align} 
		y_R = \mathbf{a}^H\mathbf g + z, \label{eq}
		\end{align}		
	\end{lemma}
	where 
	$\mathbf a=[1,\mathbf v,\mathbf v \otimes \mathbf v]^T$ and $\mathbf g=[ h_{d}^2,2 h_{d} \mathbf h_c,\mathbf h_c \otimes \mathbf h_c]^T$.
	\begin{IEEEproof}
		By expanding \eqref{y_r}, we have
		\begin{align} 
		y_{R} &=   h_{d}^2 + 2 h_{d}\mathbf v^H \mathbf h_c + \big(\mathbf v^H \mathbf h_c \big)^2  + z, \label{y_r22}
		\end{align}
		where 
		\begin{align} 
		\big(\mathbf v^H \mathbf h_c \big)^2 &= \big(\mathbf v^H \mathbf h_c \big)  \big(\mathbf v^H \mathbf h_c \big) \nonumber \\
		&=\begin{bmatrix}
		 v_1\mathbf v, \dots,v_N\mathbf v \nonumber\\
		\end{bmatrix}^H
		\begin{bmatrix}
		 h_{c,1}\mathbf h_c\\
		\vdots\\
		 h_{c,N}\mathbf h_c
		\end{bmatrix} 
		\\
		&=(\mathbf v \otimes \mathbf v)^H (\mathbf h_c \otimes \mathbf h_c),
		\end{align}
		with $v_n$ and $h_{c,n}$ being the $n$-th element of vectors $\mathbf v$ and $\mathbf h_c$, respectively. Thus, \eqref{y_r22} can be equivalently expressed as
		\begin{align} 
		y_{R} &=  h_{d}^2 + 2 h_{d}\mathbf v^H \mathbf h_c + (\mathbf v \otimes \mathbf v)^H (\mathbf h_c \otimes \mathbf h_c)   + z \nonumber \\
		&= \mathbf a^H \mathbf g +z. \nonumber 
		\end{align}
		The proof is thus completed.
	\end{IEEEproof}	
	Note that in \eqref{eq}, $\mathbf a$ depends on  the IRS training reflection $\mathbf v$ only and $\mathbf g$ is determined by the CSI (i.e., $ h_{d}$ and $\mathbf h_c$) only.
		
	\begin{lemma} \label{lemma1}
		The optimal IRS passive beamforming for data transmission  in \eqref{opt} can be equivalently rewritten as
		\begin{align} 
		\mathbf v^\ast = e^{\jmath \arg \big({ {( h^2_d)}^\dag}  (2 h_d \mathbf h_c) \big)} = e^{\jmath\arg\big( { { g}}_1^\dag [{\mathbf g}]_{2:N+1}\big)}, \label{opt2}
		\end{align}	
		where $[g_1,\dots,g_{N^2+N+1}]^T=\mathbf g$.	
	\end{lemma}
	\begin{IEEEproof}
		First, we have $\arg(  { h}_d^\dag \mathbf h_c) =\arg( | h_d| { h}_d^\dag \mathbf h_c)= \arg(  h_d { h}_d^\dag { h}_d^\dag \mathbf h_c)=  \arg \Big({ ({ h}_d^2)}^\dag   h_d \mathbf h_c\Big) ={ \arg \Big({ {( h^2_d)}^\dag}  (2 h_d \mathbf h_c) \Big)}$. Second, it follows  from \eqref{eq} that ${ g}_1= h_d^2$ and $[{\mathbf g}]_{2:N+1} = 2 h_d \mathbf h_c$. Combining the above leads to the desired result.
	\end{IEEEproof}

	Lemma \ref{lemma1} shows that it is sufficient to acquire the CSI of the first $N+1$ elements of $\mathbf g$ (i.e., $ h_d^2$ and $2 h_{d} \mathbf h_c$) for designing the optimal IRS passive beamforming for data transmission. Thus, we propose an efficient channel estimation scheme in the next  to estimate $ g_1$ and $[{\mathbf g}]_{2:N+1}$. 
	
	
	Our key idea is by leveraging the \emph{phase-rotated} IRS training reflections to resolve the sign ambiguity issue. Specifically, the proposed channel estimation scheme consists of $K$ sub-blocks. In each sub-block $k$, the reader sends two pilot signals while the IRS sets its training reflections as  $\mathbf v_k=[v_{k,1},\dots,v_{k,N}]^T$ and its phase-rotated version  $e^{\jmath\varphi}\mathbf v_k$, with $\varphi\in [0,2\pi)$ over the two symbols. Let $y_{k}^{(1)}$ and $y_{k}^{(2)}$ denote the two received pilot signals in sub-block $k$. Based on \eqref{y_r22}, we  have 
	\begin{align} 
	y_{k}^{(1)} &=  h_{d}^2 + 2 h_{d}\mathbf v_{k}^H \mathbf h_c + w_k  + z_{k_1}, \label{y1} \\
	y_{k}^{(2)} &=   h_{d}^2 + 2e^{\jmath\varphi} h_{d}\mathbf v_{k}^H \mathbf h_c + e^{2\jmath\varphi}w_k  + z_{k_2}, \label{y2}
	\end{align}
	where $ w_k=(\mathbf v_{k} \otimes \mathbf v_{k})^H (\mathbf h_c \otimes \mathbf h_c)$.
	Let $ t_1 = e^{2\jmath\varphi}$, $ t_2 = (e^{2\jmath\varphi}-1)$,  and $ t_3 = (e^{2\jmath\varphi}-e^{\jmath\varphi})$. 
	It then follows that 	
	\begin{align} 
	 t_1 y_{k}^{(1)}-y_{k}^{(2)} &=    t_2 h_{d}^2 + 2 t_3 h_{d}\mathbf v_{k}^H \mathbf h_c    + \underline z_{k},  \label{eq9}
	\end{align}
	where $\underline z_{k}= ( t_1 z_{k_1}-z_{k_2}) $ is the AWGN noise with (normalized) power $2\sigma^{2}/P_t$. For the received signals over the $K$ sub-blocks, by defining  $\underline {\mathbf y}_R =[ t_1 y_{1}^{(1)}-y_{1}^{(2)},\dots, t_1 y_{K}^{(1)}-y_{K}^{(2)}]^T \in \mathbb C^{K\times 1}$,
	\begin{align} 
	\underline{\mathbf A} &=\begin{bmatrix}
	 t_2 &  t_3\mathbf v_1^H  \\
	\vdots  & \vdots \\
	 t_2 &  t_3\mathbf v_K^H  
	\end{bmatrix} \in \mathbb C^{K\times (N+1)}, \label{mat_A}
	\end{align}
$ \underline {\mathbf g} =[ h_{d}^2,2 h_{d} \mathbf h_c]^T\in \mathbb C^{(N+1)\times 1}$, and $\underline{\mathbf z}=[\underline z_1,\dots,\underline z_K]^T\in \mathbb C^{K\times 1}$, 
	we have
	\begin{align}
	\underline {\mathbf y}_R &=  \underline{\mathbf A}~ \underline{\mathbf g} + \underline{\mathbf z}, \label{eq10}
	\end{align}
	where  $\underline{\mathbf A}$ is defined as the effective IRS training reflection matrix over $K$ sub-blocks  and $\underline{\mathbf g}$ is the required  CSI for the passive beamforming design. Next, by properly designing $\{\mathbf v_k\}_{k=1}^K$ and $\varphi$ (or equivalently $ t_2$ and $ t_3$) such that $\rank(\underline{\mathbf A}) = N+1$, the least-square (LS)  estimation  of $\underline{\mathbf g} $ can be obtained as 
	\begin{align} 
	\hat {\underline{\mathbf g}} &=  \big(\underline{\mathbf A}^H \underline{\mathbf A}\big)^{-1}  \underline{\mathbf A}^H \underline {\mathbf y}_R=\underline{\mathbf g}+\underline{\mathbf g}_{\mathrm e}, \label{ls}
	\end{align}
	where $\underline{\mathbf g}_{\mathrm e}\defeq \big(\underline{\mathbf A}^H \underline{\mathbf A}\big)^{-1}  \underline{\mathbf A}^H \underline{\mathbf z}$ denotes the channel estimation error in $\hat {\underline{\mathbf g}}$. 
	Note that to ensure $\rank(\underline{\mathbf A}) = N+1$, at least $N+1$ sub-blocks and hence equivalently $2(N+1)$ training symbols are required for estimating $\underline {\mathbf g}$. According to \eqref{ls}, the  mean-square error (MSE) of the above LS estimation is given by
	\begin{align} 
	\text{MSE}(\hat {\underline{\mathbf g}})&=\mathbb E \Big[\|\underline{\mathbf g}-\hat {\underline{\mathbf g}} \|^2\Big]= \mathbb E \big[\|\underline{\mathbf g}_{\mathrm e} \|^2\big] \nonumber\\
	&=\mathbb E \Big[\Tr\Big(\big(\underline{\mathbf A}^H \underline{\mathbf A}\big)^{-1}  \underline{\mathbf A}^H \underline{\mathbf z}\underline{\mathbf z}^H\underline{\mathbf A}\Big(\big(\underline{\mathbf A}^H \underline{\mathbf A}\big)^{-1}\Big)^H\Big)\Big] \nonumber\\
	&= \Big(\frac{2\sigma^2}{P_t}\Big) \Tr\Big(\big(\underline{\mathbf A}^H \underline{\mathbf A}\big)^{-1}\Big). \label{mse}
	\end{align}	
	
	\subsection{Proposed Training Design}

	In this subsection, we aim to minimize the MSE in \eqref{mse} by optimizing the effective IRS training reflection matrix $\underline{\mathbf A}$. Under the constraints on the unit-modulus elements in  $\{\mathbf v_k\}_{k=1}^K$ and full rank  of $\underline{\mathbf A}$, this optimization problem can be formulated as follows (by dropping the constant term  $2\sigma^2/P_t$).
	\begin{align} 
	\mathrm{(P1)}:  
	\mathop{\mathtt{minimize}}_{\varphi, \{\mathbf v_k\}_{k=1}^K}~~&  \Tr\Big(\big(\underline{\mathbf A}^H \underline{\mathbf A}\big)^{-1}\Big)  \label{eq:P2_Obj} \\
	\mathtt{s.t.} 
	~~& | v_{k,n}| = 1, \forall k, n = 1,\dots,N, \label{modulus} \\
	~~& \rank(\underline{\mathbf A}) = N+1, \label{rank}  
	\end{align}
	where the training reflection matrix $\underline{\mathbf A}$ is given in \eqref{mat_A}.


	\begin{figure*}[t!]
		\begin{align} 
		\underline{\mathbf A}^H \underline{\mathbf A} =  \begin{bmatrix}
		K| t_2|^2 & { t}_2^\dag t_3\sum_{k=1}^{K}{ v}_{k,1}^\dag& { t}_2^\dag t_3\sum_{k=1}^{K}{ v}_{k,2}^\dag & \dots& { t}_2^\dag t_3\sum_{k=1}^{K}{ v}_{k,N}^\dag \\
		{ t}_2 { t}_3^\dag\sum_{k=1}^{K}{ v}_{k,1} & K| t_3|^2& | t_3|^2\sum_{k=1}^{K}{ v}_{k,1}{ v}_{k,2}^\dag&\dots&| t_3|^2\sum_{k=1}^{K}{ v}_{k,1}{ v}_{k,N}^\dag\\
		\vdots  & \vdots& \vdots & \ddots & \vdots\\
		{ t}_2 { t}_3^\dag\sum_{k=1}^{K}{ v}_{k,N} & | t_3|^2\sum_{k=1}^{K}{ v}_{k,N}{ v}_{k,1}^\dag& | t_3|^2\sum_{k=1}^{K}{ v}_{k,N}{ v}_{k,2}^\dag &\dots &K| t_3|^2
		\end{bmatrix} \in \mathbb C^{(N+1)\times(N+1)}\label{eq16} \\ 
		\hline \nonumber
		\end{align}	   
	\end{figure*}

	Problem (P1) is a non-convex optimization problem  due to the matrix  inverse operation in the objective function and the  non-convex constraints in \eqref{modulus} and \eqref{rank}, and
	thus is difficult to be optimally solved in general. To address this issue, we first derive a lower bound of the objective function  and then obtain the optimal solution to problem (P1) in closed form. 
	\begin{lemma} \label{tr_pro}
		 The objective function of (P1) is lower-bounded by
		\begin{align}
		\Tr\Big(\big(\underline{\mathbf A}^H \underline{\mathbf A}\big)^{-1}\Big) \geq \frac{(N+1)}{\phi} , \label{lbound1}
		\end{align}
		where the equality is achieved if and only if   $\underline{\mathbf A}^H \underline{\mathbf A}=\phi \mathbf I $.
	\end{lemma} 
	\begin{IEEEproof}
		Let  $\underline{\mathbf B} \defeq  \big(\underline{\mathbf A}^H \underline{\mathbf A}\big)^{-1}$. According to \cite{923849},  we have 
		\begin{align}
		\Tr\Big(\big(\underline{\mathbf A}^H \underline{\mathbf A}\big)^{-1}\Big) \geq (N+1) \sqrt[N+1]{\det(\underline{\mathbf B})}, \label{lbound}
		\end{align}
		where the equality is achieved when $\underline{\mathbf B}$  is  diagonal  and  all the diagonal elements $\underline{\mathbf B}_{i,i}$, $i=1,\dots,N+1$, are equal. Note that when $\underline{\mathbf B}$ is diagonal and $\underline{\mathbf B}_{i,i}=\frac{1}{\phi},\forall i$, with  $\phi>0$,  the matrix $\underline{\mathbf A}^H \underline{\mathbf A}$ is also diagonal with all diagonal elements being $\phi$. Then we have $\det(\underline{\mathbf B})=\frac{1}{\det(\underline{\mathbf A}^H \underline{\mathbf A})}=\frac{1}{\phi^{N+1}}$. Substituting $\det(\underline{\mathbf B})$ into \eqref{lbound} leads to the lower bound given in \eqref{lbound1}.  The proof is thus completed.
	\end{IEEEproof}

	Next, the optimal solution to problem (P1) is given in the following proposition, as it  achieves the lower bound given in \eqref{lbound1}.
	\begin{proposition} \label{tr_pro2}
		The optimal solution to problem (P1) for minimizing the MSE in \eqref{mse} should satisfy:
		\begin{enumerate}  
			\item The IRS training reflection vectors over different symbols are orthogonal, i.e.,  
			\begin{align}
			\hspace{-4mm}\sum_{k=1}^{K} {{v}^\dag_{k,i}} {v}_{k,j} =
			\begin{cases} 
			K, &  i=j, \\
			0,  &  i\neq j, \forall i,j \in \{1,\dots,N\},
			\end{cases} \label{con2} 
			\end{align}
			with each entry satisfying the unit-modulus constraint. Moreover, for each subsurface $n$, its sum of the phase shifts over $K$ symbols is zero, i.e.,
			\begin{align}
			\sum_{k=1}^{K}{v}_{k,n} &=  0, n = 1,\dots,N. \label{con1}
			\end{align}
			\item The optimal common rotated phase-shift is $\varphi^\ast = \frac{2\pi}{3}$.
		\end{enumerate}
	\end{proposition} 
	\begin{IEEEproof}
		With $\underline{\mathbf A}$ in \eqref{mat_A}, $\underline{\mathbf A}^H \underline{\mathbf A}$ can be expressed as in \eqref{eq16} (shown at the top of this page), where $| t_2|^2=2\big(1-\cos(2\varphi)\big)$ and $| t_3|^2=2\big(1-\cos(\varphi)\big)$. Since all  diagonal elements of $\underline{\mathbf A}^H \underline{\mathbf A}$ should take the same value $\phi$ (according to Lemma \ref{tr_pro}), we have
		\begin{align} 
		K| t_2|^2 &= K| t_3|^2 =\phi, \label{33} \\
		2K\big(1-\cos(2\varphi)\big) &= 2K\big(1-\cos(\varphi)\big) =\phi. \label{333}
		\end{align}
		Combining \eqref{33} and \eqref{333} leads to $\varphi_1=0~(\text{equivalently }\phi=0)$ and $\varphi_2=\frac{2\pi}{3}~(\text{equivalently }\phi=3K)$.
		Since $\phi>0$, the optimal  $\varphi$ is given by 
		\begin{align}
		\varphi^\ast&={2\pi}/{3}. 
		\end{align}
		Moreover, it is not difficult to observe from \eqref{eq16} that    
		$\underline{\mathbf A}^H \underline{\mathbf A}$ becomes  diagonal  when the conditions  in \eqref{con1} and \eqref{con2} are satisfied. The proof is thus completed.
	\end{IEEEproof}

	It is noted from Proposition \ref{tr_pro2} that we need to  construct $\{\mathbf v_k\}_{k=1}^K$ to satisfy all the conditions in \eqref{con2} and \eqref{con1}, which, however, is not a trivial task. Fortunately, we notice that the discrete Fourier transform (DFT) matrix has the perfect orthogonality  and the summation of each column (except the first column with all-one elements) is $0$, and thus it can be used  to design the optimal  IRS training reflection vectors, $\{\mathbf v_k\}_{k=1}^K$. For instance, we can construct $\{\mathbf v_k\}_{k=1}^K$ that satisfy all the conditions in \eqref{con2} and \eqref{con1}, by selecting $N(< K)$ columns from the $K\times K$ DFT matrix as
	\begin{align} 
	 v_{k,n} &= e^{-\jmath\frac{2\pi(k-1)n}{K}}, \forall k\in\{1,\dots,K\};~n=1,\dots,N.
	\end{align}
	
	On the other hand, the optimal  $\varphi$ in Proposition \ref{tr_pro2} can be explained as follows. By substituting $\varphi =2\pi/3$ into  $y_{k}^{(1)}$ and $y_{k}^{(2)}$ in \eqref{y1} and \eqref{y2}, respectively,  and stacking them,  we have   
	\begin{align} 
	{\begin{bmatrix}
	y_{k}^{(1)}  \\
	y_{k}^{(2)} \end{bmatrix}} = \underbrace{\begin{bmatrix}
	1 & 1 & 1 \\
	1  & e^{\jmath\frac{2\pi}{3}}  & e^{\jmath\frac{4\pi}{3}}  
	\end{bmatrix}}_{\mathbf B} \begin{bmatrix}
	 h_{d}^2  \\
	2 h_{d}\mathbf v_k^H\mathbf h_c \\
	w_k \end{bmatrix} +\begin{bmatrix}
	z_{k_1}  \\
	z_{k_2} \end{bmatrix}, \nonumber
	\end{align}	
	which satisfies $\mathbf B\mathbf B^H=3\mathbf I$. This implies that the two row vectors of $\mathbf B$ are orthogonal and hence the two received pilot signals in each sub-block $k$, i.e., $y_{k}^{(1)}$ and $y_{k}^{(2)}$, can always be obtained with minimum correlation so as to minimize the channel estimation error.

	\section{Simulation Results}
	
	We consider an IRS-assisted backscatter communication system that operates at a carrier frequency of $915$ MHz (EPC Gen 2 frequency), which corresponds to the reference path loss of $30$ dB at the distance of $1$ meter (m). 	A three-dimensional (3D) coordinate system  is considered as shown in Fig. \ref{sm2}, where the center points of the reader, tag, and IRS are located at $(2,0,0)$, $(2,13,0)$, and $(0,13,0.33)$ in m, respectively. The IRS with half-wavelength element spacing is divided into $N$ subsurface, each consisting of $5$ adjacent  reflecting elements  \cite{9195133}.	Rician fading channel model is assumed for all the channels involved with a Rician factor of 6 dB. The path loss exponents are set as $2.2$, $2.2$, and $3.5$ for the reader-IRS, IRS-tag, and reader-tag channels, respectively, and we set the noise power as $\sigma^2=-90 $ dBm.
	We define the system reference SNR as $10\log_{10} \big({P_t\big| h_{d}^2 \big|^2}/{\sigma^2}\big)  $ and the effective SNR for data transmission (with optimized IRS reflections applied) as $10\log_{10} \big({P_t\big|\big( h_{d} + \mathbf v^H \mathbf h_c     \big)^2 \big|^2}/{\sigma^2}\big)$, where $\mathbf v$ is designed based on the estimated channel by the proposed scheme or by other means (see the baseline schemes below).	Simulation results are averaged  over $500$ fading channel realizations. We  compare the performance of the proposed channel estimation scheme  against the following three baseline schemes: \vspace{2mm}
	
	\hspace{-0.49cm} \textbf{Baseline  I:} The channels are estimated by using the proposed scheme, while the common phase rotation  $\varphi$ is randomly set.  \vspace{1mm}\\ 
	\hspace{-0.55cm} \textbf{Baseline  II:} Randomly generate $Q$ training reflection vectors from a $Q\times Q$ DFT matrix, and select the one that results in the maximum received signal power at the reader. \vspace{1mm}\\	
	\hspace{-0.55cm} \textbf{Baseline  III:} Divide $Q$ (with $Q>N+1$) training symbols into two groups as follows. For the first group of $N+1$ symbols, denoted by the set $\Omega_{\text{1}}$, we try all the $2^{N+1}$ sign combinations for $\{b_k\}_{k=1}^{N+1}$ in \eqref{ehau} and obtain the corresponding candidate CSI by using the LS estimation method in \cite{9195133,9053695,91331422}. Then, for the remaining $Q-(N+1)$ training symbols, denoted by the set $\Omega_{\text{2}}$, we first  construct  the received signals for each candidate CSI based on the prior known  training reflections and then obtain their  MSEs with the actual received signals. Last, the candidate CSI that yields the minimum testing MSE over the second group of symbols is selected. 
	
	\begin{figure}[t!] 
		\centering 
		\includegraphics[width=0.87\linewidth]{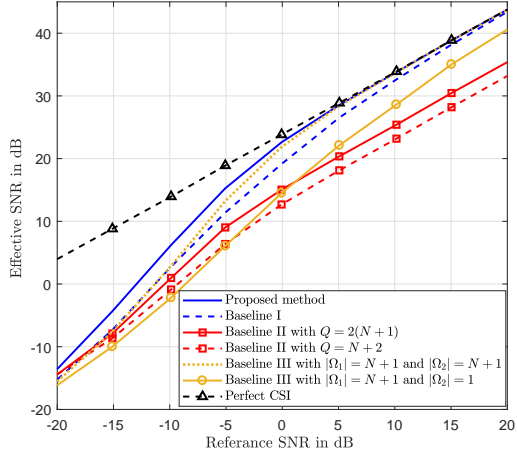}
		\caption{{Effective SNR versus reference SNR.}}  
		\label{f1} 
	\end{figure}	
	
	In Fig. \ref{f1}, we plot the effective  SNR versus reference SNR by varying the normalized noise power (i.e., $\sigma^2/P_t$) from $-120$ dB to $-160$ dB with $N=10$. First, it is observed that given the same number of $2(N+1)$ training symbols, the proposed scheme achieves smaller MSE than the other baseline schemes and approaches close performance to the scheme with perfect CSI that yields an upper bound on the effective SNR.	
	Second,  in the high SNR regime, Baseline  III with $|\Omega_2|= 1$  performs very close to the proposed method. This is due to the fact that at high SNR, it is very likely that only one candidate CSI achieves the minimum testing MSE.   However,  Baseline III in general requires exponentially increasing complexity due to the required exhaustive search of $2^{N+1}$ sign combinations, and thus is computationally prohibitive in practice. In contrast, the proposed scheme achieves superior MSE performance in both the low and high SNR regimes and requires low complexity (see \eqref{ls}).

\begin{figure}[t!]
		\centering 
		\includegraphics[width=0.87\linewidth]{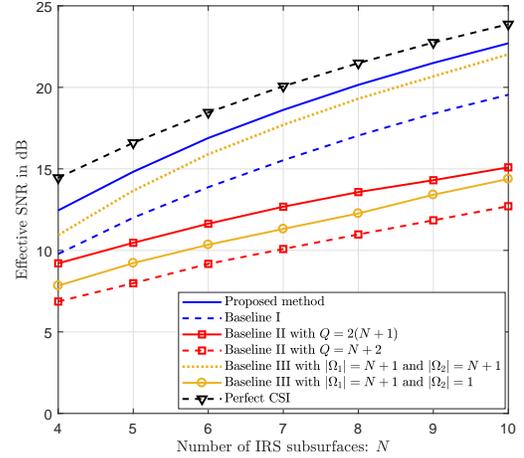}
		\caption{{Effective SNR versus number of IRS subsurfaces.}}  
		\label{f2} 
	\end{figure}

	Next, we plot in Fig. \ref{f2} the effective SNR versus the number of IRS subsurfaces  with fixed reference SNR of $0$ dB. It is observed that the proposed scheme significantly outperforms the other baseline schemes, and the  performance gap between the perfect CSI case and our proposed scheme decreases with  increasing $ N$. It is also observed that the performance of Baseline III with  $|\Omega_2|=1$ deviates away from that of  Baseline III with $|\Omega_2|= N+1$ as $N$ increases. This is because a larger $N$ requires more training symbols and hence more possible sign 	combinations that generally needs more training  symbols in $\Omega_2$  to determine the CSI with high accuracy.
	
	\section{Conclusion}  
	
	In this letter, we considered an IRS-assisted monostatic backscatter communication system, where an IRS is deployed to assist in the   communication between a full-duplex single-antenna reader and a single-antenna tag. We proposed an efficient channel estimation scheme to estimate both the reader-tag direct channel and reader-IRS-tag reflecting channel, and optimized the IRS training matrix for minimizing the channel estimation error. Simulation results  verified the superior performance of our proposed scheme over the baseline schemes. In future work, it is worth studying more general system/channel setups, such as the bistatic backscatter system \cite{jia2020intelligentxx}, multi-tag system, IRS with discrete phase shift levels \cite{IRS_discrete} or practical phase shift model \cite{9115725}.

	\bibliographystyle{IEEEtran}  
	\footnotesize{\bibliography{bibfile}}

\begin{thebibliography}{10}
\providecommand{\url}[1]{#1}
\csname url@samestyle\endcsname
\providecommand{\newblock}{\relax}
\providecommand{\bibinfo}[2]{#2}
\providecommand{\BIBentrySTDinterwordspacing}{\spaceskip=0pt\relax}
\providecommand{\BIBentryALTinterwordstretchfactor}{4}
\providecommand{\BIBentryALTinterwordspacing}{\spaceskip=\fontdimen2\font plus
\BIBentryALTinterwordstretchfactor\fontdimen3\font minus
  \fontdimen4\font\relax}
\providecommand{\BIBforeignlanguage}[2]{{%
\expandafter\ifx\csname l@#1\endcsname\relax
\typeout{** WARNING: IEEEtran.bst: No hyphenation pattern has been}%
\typeout{** loaded for the language `#1'. Using the pattern for}%
\typeout{** the default language instead.}%
\else
\language=\csname l@#1\endcsname
\fi
#2}}
\providecommand{\BIBdecl}{\relax}
\BIBdecl

\bibitem{wu2020intelligent}
Q.~{Wu}, S.~{Zhang}, B.~{Zheng}, C.~{You}, and R.~{Zhang}, ``Intelligent
  reflecting surface aided wireless communications: A tutorial,'' \emph{IEEE
  Trans. Commun., DOI:10.1109/TCOMM.2021.3051897}, Jan. 2021.

\bibitem{chongwang}
C.~{Huang}, A.~{Zappone}, G.~C. {Alexandropoulos}, M.~{Debbah}, and C.~{Yuen},
  ``Reconfigurable intelligent surfaces for energy efficiency in wireless
  communication,'' \emph{IEEE Trans. Wireless Commun.}, vol.~18, no.~8, pp.
  4157--4170, Aug. 2019.

\bibitem{9351782}
H.~{Lu}, Y.~{Zeng}, S.~{Jin}, and R.~{Zhang}, ``Aerial intelligent reflecting
  surface: Joint placement and passive beamforming design with {3D} beam
  flattening,'' \emph{IEEE Trans. Wireless Commun.,
  DOI:10.1109/TWC.2021.3056154}, Feb. 2021.

\bibitem{9279326}
F.~{Zhou}, C.~{You}, and R.~{Zhang}, ``Delay-optimal scheduling for {IRS}-aided
  mobile edge computing,'' \emph{IEEE Wireless Commun. Lett.,
  DOI:10.1109/LWC.2020.3042189}, Dec. 2020.

\bibitem{9017956}
W.~{Zhao}, G.~{Wang}, S.~{Atapattu}, T.~A. {Tsiftsis}, and X.~{Ma},
  ``Performance analysis of large intelligent surface aided backscatter
  communication systems,'' \emph{IEEE Wireless Commun. Lett.}, vol.~9, no.~7,
  pp. 962--966, Jul. 2020.

\bibitem{jia2020intelligentxx}
X.~Jia, X.~Zhou, D.~Niyato, and J.~Zhao, ``Intelligent reflecting
  surface-assisted bistatic backscatter networks: {Joint} beamforming and
  reflection design,'' [Online]. Available: https://arxiv.org/abs/2010.08947.

\bibitem{8618337}
D.~{Mishra} and E.~G. {Larsson}, ``Optimal channel estimation for
  reciprocity-based backscattering with a full-duplex {MIMO} reader,''
  \emph{IEEE Trans. Signal Process.}, vol.~67, no.~6, pp. 1662--1677, Mar.
  2019.

\bibitem{91331422}
C.~{You}, B.~{Zheng}, and R.~{Zhang}, ``Channel estimation and passive
  beamforming for intelligent reflecting surface: Discrete phase shift and
  progressive refinement,'' \emph{IEEE J. Sel. Areas Commun.}, vol.~38, no.~11,
  pp. 2604--2620, Nov. 2020.

\bibitem{9195133}
B.~{Zheng}, C.~{You}, and R.~{Zhang}, ``Intelligent reflecting surface assisted
  multi-user {OFDMA}: Channel estimation and training design,'' \emph{IEEE
  Trans. Wireless Commun.}, vol.~19, no.~12, pp. 8315--8329, Dec. 2020.

\bibitem{9053695}
T.~L. {Jensen} and E.~{De Carvalho}, ``An optimal channel estimation scheme for
  intelligent reflecting surfaces based on a minimum variance unbiased
  estimator,'' in \emph{Proc. ICASSP}, May 2020, pp. 5000--5004.

\bibitem{9129778}
C.~You, B.~Zheng, and R.~Zhang, ``Fast beam training for {IRS}-assisted
  multiuser communications,'' \emph{IEEE Wireless Commun. Lett.}, vol.~9,
  no.~11, pp. 1845--1849, Nov. 2020.

\bibitem{chongwangxxxx}
L.~{Wei}, C.~{Huang}, G.~C. {Alexandropoulos}, C.~{Yuen}, Z.~{Zhang}, and
  M.~{Debbah}, ``Channel estimation for {RIS}-empowered multi-user {MISO}
  wireless communications,'' \emph{to appear in IEEE Trans. Commun.}, 2021.

\bibitem{5467234}
D.~P. {Villame} and J.~S. {Marciano}, ``Carrier suppression locked loop
  mechanism for {UHF} {RFID} readers,'' in \emph{Proc. IEEE Int. Conf. on
  RFID}, 2010, pp. 141--145.

\bibitem{923849}
T.~L. Tung, K.~Yao, and R.~E. Hudson, ``Channel estimation and adaptive power
  allocation for performance and capacity improvement of multiple-antenna
  {OFDM} systems,'' in \emph{Proc. IEEE SPAWC}, Mar. 2001, pp. 82--85.

\bibitem{IRS_discrete}
Q.~{Wu} and R.~{Zhang}, ``Beamforming optimization for wireless network aided
  by intelligent reflecting surface with discrete phase shifts,'' \emph{IEEE
  Trans. Commun.}, vol.~68, no.~3, pp. 1838--1851, Mar. 2020.

\bibitem{9115725}
S.~{Abeywickrama}, R.~{Zhang}, Q.~{Wu}, and C.~{Yuen}, ``Intelligent reflecting
  surface: Practical phase shift model and beamforming optimization,''
  \emph{IEEE Trans. Commun.}, vol.~68, no.~9, pp. 5849--5863, Sep. 2020.

\end{thebibliography}
	
\end{document}